\documentclass[aps,twocolumn,showpacs,superscriptaddress,amssymb]{revtex4}
\date{\today}

\usepackage{graphicx,amssymb}

\begin{document}

\title{Deformation and its influence on $K$ isomerism in neutron-rich Hf nuclei}

\author{H.L. Liu}
\affiliation{Department of Physics and Astronomy, Texas A\&M
University-Commerce, Commerce, Texas 75429-3011, USA}
\author{F.R. Xu}
\affiliation{School of Physics, Peking University, Beijing 100871,
China}
\author{P.M. Walker}
\affiliation{Department of Physics, University of Surrey,
Guildford, Surrey GU2 7XH, UK}
\author{C.A. Bertulani}
\affiliation{Department of Physics and Astronomy, Texas A\&M
University-Commerce, Commerce, Texas 75429-3011, USA}

\begin{abstract}
We investigate the influence of deformation on the possible
occurrence of long-lived $K$ isomers in Hf isotopes around
$N=116$, using configuration-constrained calculations of
potential-energy surfaces. Despite having reduced shape
elongation, the multi-quasiparticle states in $^{186,188}$Hf
remain moderately robust against triaxial distortion, supporting
the long expected occurrence of exceptionally long-lived isomers.
The calculations are compared with available experimental data.
\end{abstract}

\pacs{21.10.-k, 21.60.-n, 23.20.Lv, 23.35.+g}

\maketitle

%\section{Introduction}
Atomic nuclei have special stability at some proton or neutron
numbers named magic numbers. Analogously, there exist some
$K$-magic ($K$ is the total angular momentum projection onto the
symmetry axis of a deformed nucleus) numbers associated with
exceptionally long-lived high-$K$
isomers~\cite{WalNature99,WalHI01}. Such isomers are formed
through broken-pair or multi-quasiparticle (multi-qp) excitations.
The long half-life usually originates from a combination of
relatively-low excitation energy, high $K$ value and well-deformed
axially-symmetric shape. These factors favor the conservation of
the $K$ quantum number and hence the hindrance of decay to low-$K$
states. The best known $K$-magic number is probably $Z=72$ where
$K^\pi=8^-$ isomers with the configuration
$\pi^2\{7/2^+[404]\otimes9/2^-[514]\}$ systematically occur in
$^{170-184}$Hf, with half-lives ranging from nanoseconds to
hours~\cite{WalPS83,ReedPRL10}. The neutron number $N=106$ is
another well-known $K$-magic number. Here $T_{1/2}>$1-$\mu$s
$K^\pi=8^-$ isomers with the configuration
$\nu^2\{7/2^-[514]\otimes9/2^+[624]\}$ are observed from
$^{174}$Er to $^{188}$Pb~\cite{DraPLB06}. These scarce long chains
of isomers provide unique opportunities for systematic study. The
doubly $K$-magic nucleus $^{178}$Hf manifests itself with the
occurrence of a 31-yr $K^\pi=16^+$ isomer based on the coupling
$\nu^2_{8^-}\otimes\pi^2_{8^-}$. The possible stimulated release
of the stored energy (2.4 MeV) in the isomer has been associated
with proposals for a clean reservoir of nuclear energy and
$\gamma$-ray lasers~\cite{WalNature99,WalHI01,WalPT05}.

Exceptionally long-lived isomers have been predicted to also occur
in neutron-rich Hf isotopes due to favorable conditions, i.e., low
energy and high $K$ value~\cite{WalHI01,JainNPA95,XuPRC00}. With a
$K^\pi=18^+$ isomeric state from the configuration
$\nu^2_{10^-}\{9/2^-[505]\otimes11/2^+[615]\}\otimes\pi^2_{8^-}$,
$^{188}$Hf seems to be a doubly $K$-magic nucleus~\cite{WalHI01}.
Experimentally, projectile-fragmentation~\cite{PodPLB00} and
deep-inelastic~\cite{LaPRC10} reactions have been employed to
explore this region, such as the observation of a 240-$\mu$s
isomer with the configuration $\nu^2_{10^-}$ in
$^{190}$W~\cite{PodPLB00,LaPRC10}. A major advance was achieved
recently using storage-ring mass measurements~\cite{ReedPRL10},
where long-lived isomers were observed in neutron-rich Hf isotopes
up to $^{186}$Hf, with $N=114$. Nevertheless, the predicted doubly
$K$-magic character of $^{188}$Hf, with $N=116$, remains to be
confirmed.

However, another important factor, deformation, for the emergence
of $K$ isomers could be much different between $^{178}$Hf and the
nuclei around $^{188}$Hf. With neutron number $N=106$ at the
mid-shell between spherical magic numbers 82 and 126, $^{178}$Hf
is well deformed, having a shape that favors the formation of
long-lived isomers. In contrast, a reduced shape elongation is
expected in $^{188}$Hf because the neutron number $N=116$ is far
away from the mid-shell. Generally, a smaller $\beta_2$
deformation is more susceptible to shape fluctuation towards
triaxiality that induces $K$ mixing. It is still an open question
to what extent the deformation influences the occurrence of
exceptionally long-lived $K$-isomers in neutron-rich Hf isotopes.
In the present work, we investigate these high-$K$ isomeric states
using configuration-constrained calculations of potential-energy
surfaces (PES)~\cite{XuPLB98,LiuPRC11}, a model that properly
treats the deformations of multi-qp states.

%\section{The Model}
In our model, the single-particle levels are obtained from the
deformed Woods-Saxon potential with the universal parameter
set~\cite{NazNPA85,CwCPC87}. To avoid the spurious pairing phase
transition encountered in the usual BCS approach, we use the
Lipkin-Nogami (LN) treatment of pairing~\cite{LipNPA73}, with
pairing strength $G$ determined by the average-gap
method~\cite{MolNPA92}. The configuration energy~\cite{XuPLB98} in
the LN approach can be written as
\begin{eqnarray}
E_{\text{LN}} & = & \sum_{j=1}^{S}e_{k_j}+\sum_{k\neq k_j}2V_k^2e_k
-\frac{\Delta^2}{G}-G\sum_{k\neq k_j}V_k^4 \nonumber \\
&&+G\frac{N-S}{2}-4\lambda_2\sum_{k\neq k_j}(U_kV_k)^2,
\end{eqnarray}
where $S$ is the proton or neutron seniority for a specified
configuration (i.e., the number of blocked orbits with index
$k_j$), and $N$ is the proton or neutron number. The orbits with
index $k_j$ are blocked at each point of the
($\beta_2,\gamma,\beta_4$)~\cite{XuPLB98} or
($\beta_2,\beta_4,\beta_6$)~\cite{LiuPRC11} deformation lattice,
which is the so-called configuration constraint. This is achieved
by calculating and identifying the average Nilsson quantum numbers
for every orbit involved in a configuration. The total energy of a
configuration consists of a macroscopic part which is obtained
from the standard liquid-drop model~\cite{MyeNP66} and a
microscopic part resulting from the Strutinsky shell
correction~\cite{StruNPA67}, $\delta
E_{\text{shell}}=E_{\text{LN}}-\tilde{E}_{\text{Strut}}$. The
minimum of the PES gives the energy, deformation and pairing
property of a multi-qp state.

%\section{Calculations}
\begin{figure}
\includegraphics[scale=0.54]{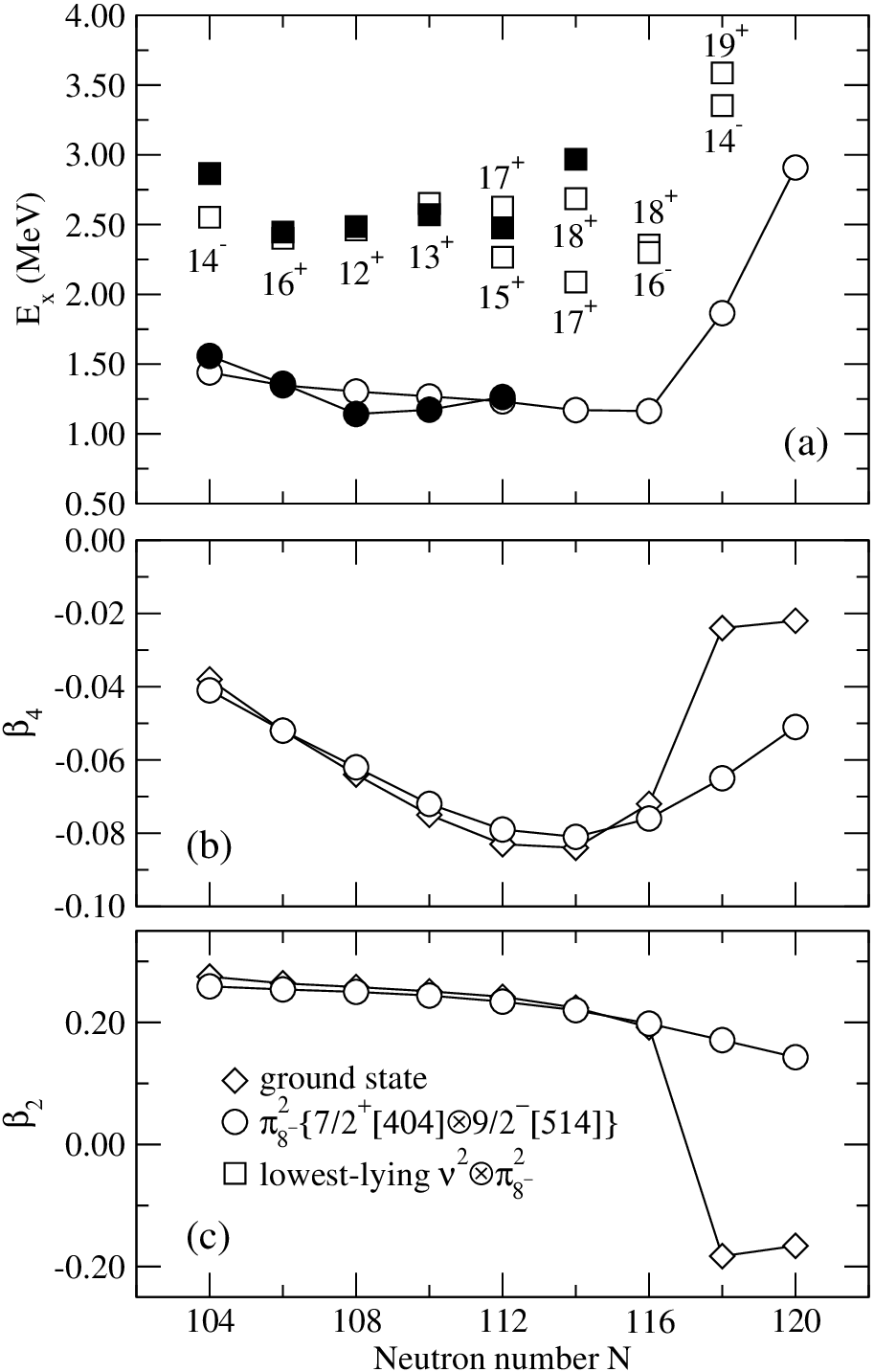}
\caption{\label{fig1}Excitation energies (a), hexadecapole
deformations (b), and quadrupole deformations (c) for multi-qp
isomeric states in Hf isotopes. Open and filled symbols show the
calculated and experimental
values~\cite{ReedPRL10,DraPLB06,AudNPA03}, respectively. The data
for the 2-qp state in $^{178}$Hf ($N=106$) corresponds to an
unperturbed $\pi^2_{8^-}$ state (see text). Note that no
experimental configuration is available for the nuclei with
$N\ge112$. The deformations of 4-qp states (not shown for clarity)
are similar to those of the corresponding 2-qp states. For all the
multi-qp states, $\gamma\approx0$ and $\beta_6\approx0$. The same
configurations are connected by lines to guide the eye. See
Table~\ref{tab1} for the neutron configurations $\nu^2$ of the
4-qp isomeric states.}
\end{figure}

Figure~\ref{fig1} displays the calculated excitation energies and
deformations of high-$K$ isomeric states in even-$A$
$^{176-192}$Hf isotopes. Our calculations satisfactorily reproduce
the excitation energies of the $\pi^2$ $8^-$ isomers observed in
$^{176-182}$Hf~\cite{AudNPA03}. It should be noted that the
observed 4-s isomer in $^{178}$Hf has a mixed configuration of
64\% $\nu^2_{8^-}$ and 36\% $\pi^2_{8^-}$~\cite{MuPLB97},
resulting from the interaction between the two close-lying
$\nu^2_{8^-}$ and $\pi^2_{8^-}$ configurations. The experimental
excitation energy of the unperturbed $\pi^2_{8^-}$ state, 1.35
MeV~\cite{DraPLB06}, is well reproduced by our calculation of 1.35
MeV. The calculation of the $8^-$ state in $^{184}$Hf gives an
excitation energy of 1.23 MeV that is in good agreement with the
measured value of 1.26 MeV~\cite{ReedPRL10}. This indicates the
association of the isomer with the $\pi^2_{8^-}$ configuration,
which is consistent with the suggestion in Ref.~\cite{ReedPRL10}.

The calculated excitation energy of the $K^\pi=8^-$ state suddenly
increases at $N=118$, as shown in Fig.~\ref{fig1}. Here there
occurs a ground-state phase transition from prolate shape to
oblate shape. Our prediction is consistent with other calculations
(see, e.g., Refs.~\cite{PauPRC05,SarPRC08}). In contrast,
M{\"o}ller {\it et al.} predict that $^{190}$Hf is still prolate
deformed~\cite{Mo95}, indicating that the prediction of the
neutron number for the prolate-to-oblate transition is model
dependent. The $8^-$ state in $^{190}$Hf is calculated (see
Fig.~\ref{fig1}) to be at an excitation energy of 1.86 MeV, but
has a prolate shape, distinct from the oblate shape of the ground
state. Also the calculated $\beta_4$ deformations are much
different between the isomeric state and the ground state. A
combination of $K$ isomerism and shape isomerism may happen in
$^{190}$Hf, analogous to that predicted for the $\nu^2$ $8^-$
isomer in $^{186}$Hg~\cite{XuPRC99}. Whether or not the higher
excitation energies and smaller $\beta_2$ deformations predicted
for the more neutron-rich Hf isotopes will lead to the termination
of the $K^\pi=8^-$ isomer chain at $N\approx118$ remains to be
determined.

\begin{table}
\caption{\label{tab1}Neutron configurations of the
$\nu^2\otimes\pi^2_{8^-}\{7/2^+[404]\otimes9/2^-[514]\}$ isomeric
states.}
\begin{ruledtabular}
\begin{tabular}{cccc}
Nucleus & $K^\pi_{\text{4-qp}}$ & $K^\pi_{\nu^2}$ & Neutron configuration $\nu^2$ \\
\hline
 $^{176}$Hf$_{104}$ & $14^-$ & $6^+$ & $\ 5/2^-[512]\otimes\ 7/2^-[514]$ \\
 $^{178}$Hf$_{106}$ & $16^+$ & $8^-$ & $\ 7/2^-[514]\otimes\ 9/2^+[624]$ \\
 $^{180}$Hf$_{108}$ & $12^+$ & $4^-$ & $\ 9/2^+[624]\otimes\ 1/2^-[510]$ \\
 $^{182}$Hf$_{110}$ & $13^+$ & $5^-$ & $\ 1/2^-[510]\otimes11/2^+[615]$ \\
 $^{184}$Hf$_{112}$ & $15^+$ & $7^-$ & $\ 3/2^-[512]\otimes11/2^+[615]$ \\
                    & $17^+$ & $9^-$ & $11/2^+[615]\otimes\ 7/2^-[503]$ \\
 $^{186}$Hf$_{114}$ & $17^+$ & $9^-$ & $11/2^+[615]\otimes\ 7/2^-[503]$ \\
                    & $18^+$ & $10^-$ & $11/2^+[615]\otimes\ 9/2^-[505]$ \\
 $^{188}$Hf$_{116}$ & $18^+$ & $10^-$ & $11/2^+[615]\otimes\ 9/2^-[505]$ \\
                    & $16^-$ & $8^+$ & $\ 7/2^-[503]\otimes\ 9/2^-[505]$ \\
 $^{190}$Hf$_{118}$ & $14^-$ & $6^+$ & $\ 9/2^-[505]\otimes\ 3/2^-[501]$ \\
                    & $19^+$ & $11^-$ & $\ 9/2^-[505]\otimes13/2^+[606]$ \\
\end{tabular}
\end{ruledtabular}
\end{table}

Figure~\ref{fig1} also shows the lowest-lying 4-qp high-$K$
isomeric states in $^{176-190}$Hf isotopes that are formed through
the coupling of $\nu^2$ (see Table~\ref{tab1}) with $\pi^2_{8^-}$.
For $^{188}$Hf, the $K^\pi=18^+$ state has calculated excitation
energy close to that of the $K^\pi=16^-$ state. The calculated
excitation energies are reasonably in accord with experimental
data of $^{176-182}$Hf. The observed 2.48-MeV isomer in $^{184}$Hf
is likely the lowest-lying 4-qp $15^+$ state. The calculated
excitation energy of the $17^+$ state is closer to the
experimental data, but it can barely be long-lived because of the
possibility to decay to the $15^+$ state. The newly observed
isomer in $^{186}$Hf was tentatively associated with the $17^+$
configuration in Ref.~\cite{ReedPRL10} based on blocked-Nilsson+LN
calculations with fixed deformations. That significantly
underestimated the measured excitation energy. In our
calculations, the inclusion of deformation effects does not remove
this discrepancy.

\begin{figure}
\includegraphics[scale=0.52]{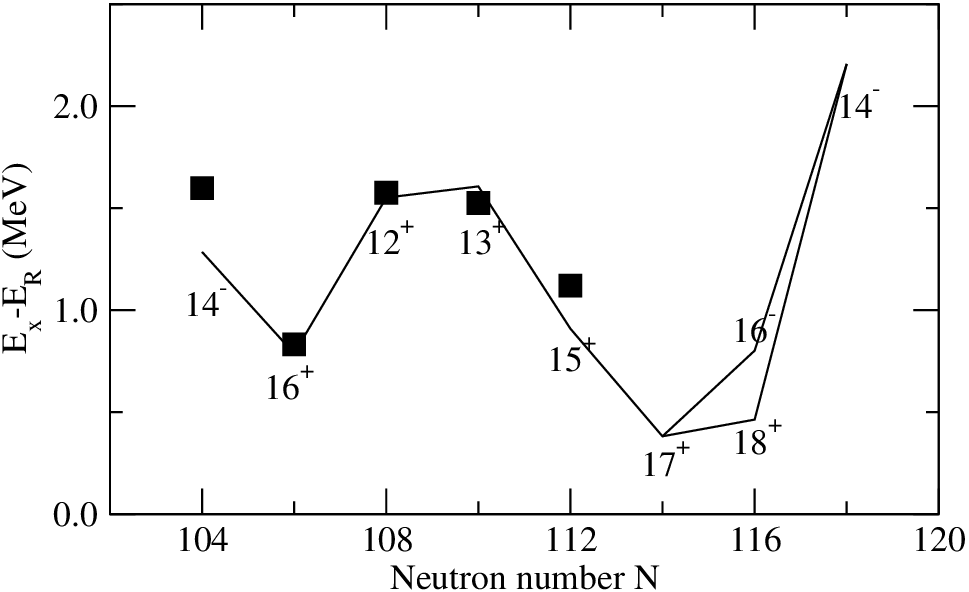}
\caption{\label{fig2}Excitation energy relative to a rigid rotor
for the lowest-lying 4-qp isomeric states in even-$A$ Hf isotopes.
Calculated and experimental values~\cite{ReedPRL10,AudNPA03} are
represented by lines and squares, respectively. The data point for
$^{186}$Hf is not shown as the $K$ value is unclear. Note that the
$16^-$ state in $^{188}$Hf has larger $E_x-E_R$ value than the
$18^+$ state because of the smaller angular momentum.}
\end{figure}

The discrepancy in $^{186}$Hf could lie in both experiment and
theory. Experimentally, the data for the 4-qp isomer in
$^{186}$Hf, obtained from just two ions, need to be confirmed.
Furthermore, the mass measurements do not give a complete picture
of the low-lying intrinsic states and do not provide spectroscopic
information. Theoretically, our calculations do not include the
residual spin-spin interaction~\cite{JainNPA95} that can change
the calculated excitation energy by up to several hundred keV. (We
note that all the neutron configurations listed in
Table~\ref{tab1} have energetically favored spin-antiparallel
couplings except for the $17^+$ state in $^{186}$Hf.) In addition,
the Woods-Saxon single-particle levels could become less accurate
when moving away from the stability line. It has been remarked
previously that the observed 1-qp states in odd-$A$ W nuclei
approaching $N=114$ begin to deviate considerably from the
predictions of the Nilsson model~\cite{JainRMP90}. Our model
cannot generate the correct state orderings in $^{187}$W either,
which implies a decreasing validity of the present Woods-Saxon
potential parameters in the neutron-rich nuclei. These parameters
are determined by fitting the properties of nuclei near the
stability line. Finally, the pairing strength could be another
factor influencing the calculated excitation energy. An
enhancement of about 10\% (relative to the value obtained from the
average-gap method) for both protons and neutrons is obtained for
$^{190}$W by fitting experimental odd-even mass
differences~\cite{WalPLB06}. This increment in pairing strength
can result in an increase of the calculated excitation energy by
$\approx400$ keV for the $\nu^2_{10^-}$ state. Unfortunately, such
a fitting process is not available for the neutron-rich Hf nuclei
due to the absence of mass data required to calculate the odd-even
mass differences. The above-mentioned problems in residual
interaction, single-particle energy and pairing strength still
await to be properly treated to obtain a more complete and
accurate description of the neutron-rich Hf nuclei.

In Fig.~\ref{fig2}, we plot the excitation energy relative to a
rigid rotor for the 4-qp isomeric states. It has been found that
the degree of $K$ hindrance increases in a simple manner with
decreasing excitation energy relative to a rigid
rotor~\cite{WalNature99,WalHI01,WalPLB97}. This is because a
relatively-low excitation energy is often associated with low
statistical $K$ mixing due to the low level
density~\cite{WalPLB97}, favoring the conservation of the $K$
quantum number. Figure~\ref{fig2} shows two minima at $N=106$ and
$N\approx116$, which is mainly ascribed to the closeness of the
two single-particle orbits occupied by the unpaired neutrons and
coupled to high $K$. The minimum at $N\approx116$ is even lower
than that at the $K$-magic number $N=106$ where the well-known
31-yr $K^\pi=16^+$ isomer occurs. This, although consistent with
the prediction in Ref.~\cite{WalHI01}, needs to be confirmed by
further work because of the uncertainty in our calculations of
excitation energy around $N=116$ (see above). It is the remarkably
low energy together with the high $K$ value that indicates the
possible emergence of exceptionally long-lived
isomers~\cite{WalHI01,JainNPA95,XuPRC00}.

\begin{figure}
\includegraphics[scale=0.52]{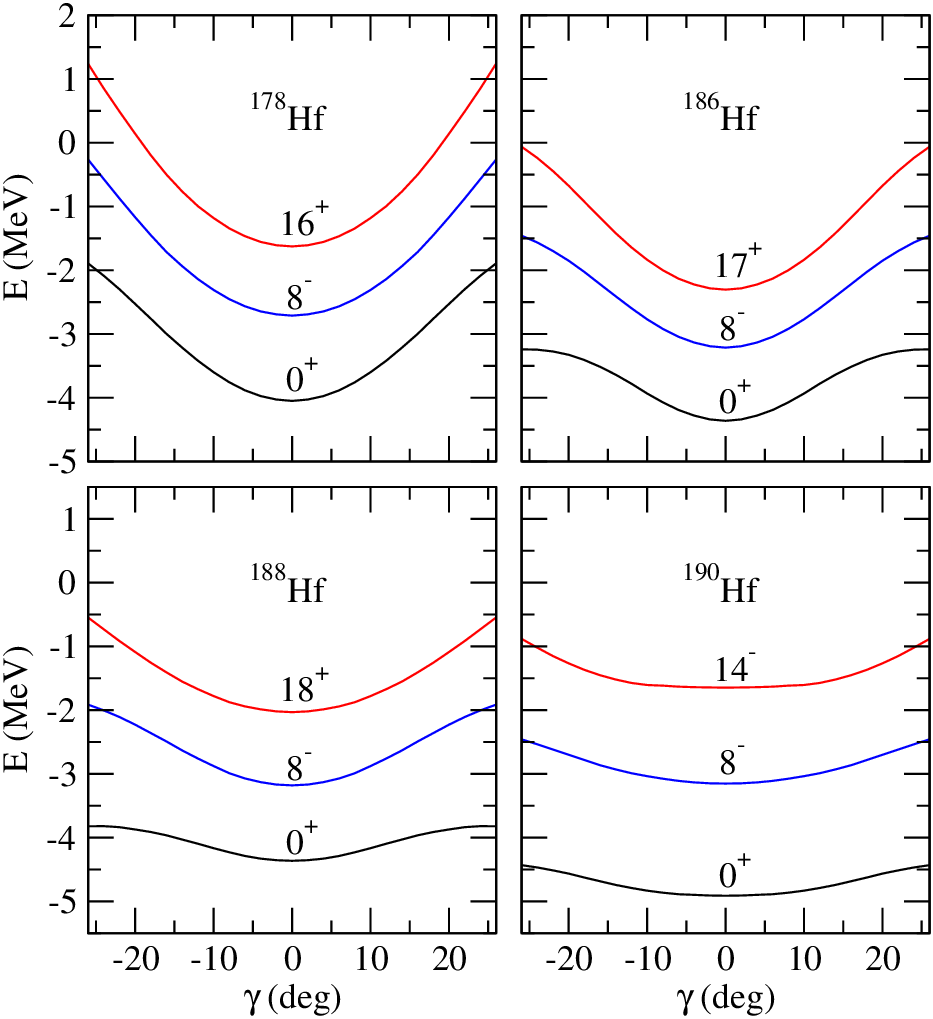}
\caption{\label{fig3}(Color online)Calculated potential-energy
curves as a function of $\gamma$ deformation. The energy is
minimized with respect to $\beta_2$ and $\beta_4$ deformations.
The true minimum of the oblate-deformed ground state in $^{190}$Hf
is at $\gamma=-60^\circ$ in the Lund convention.}
\end{figure}

Nevertheless, these isomeric states have smaller $\beta_2$
deformations than $^{178}$Hf. It is usually easier for a less
elongated shape to fluctuate towards $\gamma$ distortion. Axially
asymmetric deformation is intimately involved with $K$ mixing,
because there is then no axis of symmetry on which to define a $K$
projection. Some high-$K$ isomers with very low hindrance in decay
have been found to have $\gamma$ deformations (see, e.g.,
Ref.~\cite{XuPLB98}). Even though a $K$ isomer has axially
symmetric shape, it can decay by tunnelling through triaxial
shapes~\cite{NarNPA96}. The softness against $\gamma$ deformation
facilitates such a decay mode. To see the degree of $\gamma$
softness accompanied by small $\beta_2$ deformation, we calculate
the potential-energy curves as a function of $\gamma$ deformation
for Hf isotopes, as shown in Fig.~\ref{fig3}. The doubly $K$-magic
nucleus $^{178}$Hf can be treated as a benchmark, to which the
neutron-rich nuclei are compared. The qualitative comparison can
be barely influenced by the potential inaccuracy in pairing
strengths and single-particle levels. It can be seen that the
ground states of neutron-rich Hf isotopes become increasingly soft
against triaxial distortion with decreasing $\beta_2$
deformations. With $|\beta_2|<0.2$, $^{188,190}$Hf have remarkably
flat ground-state potential wells. However, the multi-qp states in
$^{186,188}$Hf show enhanced stiffness in the potential wells
compared to the corresponding ground states. The isomeric states
in $^{188}$Hf have potential wells not much different from those
in $^{186}$Hf where a long-lived isomer has been
observed~\cite{ReedPRL10}, though the depth of the potential wells
in both are moderately reduced with respect to $^{178}$Hf. This,
together with the favorable conditions in energy and $K$ value,
points to the possibility of very long-lived isomers in
$^{186,188}$Hf.

The neutron-rich Hf isomeric states are expected to decay through
$\beta^-$ emission competing with $\gamma$-ray
transitions~\cite{WalHI01,ReedPRL10}. The $K^\pi=8^-$ isomers in
$^{178,180}$Hf mainly decay to the $8^+$ member in the
ground-state band by $K$-forbidden $E1$
transitions~\cite{AudNPA03}. For neutron-rich nuclei, the $8^+$
state would lie higher in energy than the isomer due to reduced
$\beta_2$ deformation. Instead, the isomers would decay to the
$6^+$ member by slower $M2/E3$ transitions, making the
$\beta^-$-decay channel competitive. A similar situation could
happen for the 4-qp isomeric states where the major
$\gamma$-transition channel is open to the members in the
rotational band built upon the $K^\pi=8^-$ isomer. Spectroscopic
measurements may soon be available owing to the quick development
of experimental techniques.

%\section{Summary}
In summary, the multi-qp states in neutron-rich Hf isotopes are
investigated by using configuration-constrained PES calculations,
with attention paid to see the influence of deformations on the
occurrence of long-lived isomers. Isomeric states with
relatively-low excitation energy and very high $K$ value are
predicted in Hf isotopes at $N\approx116$. In spite of having
small $\beta_2$ deformations, the multi-qp states in
$^{186,188}$Hf remain moderately robust against $\gamma$
distortion, favoring the formation of long-lived isomers. Further
work in both experiment and theory is needed to study the isomeric
structures in neutron-rich Hf isotopes, through the $N\approx116$
prolate-oblate shape-transition region.

%\acknowledgments
This work was supported in part by the US DOE under Grants No.
DE-FG02-08ER41533 and No. DE-FC02-07ER41457 (UNEDF, SciDAC-2), and
the Research Corporation; the Chinese Major State Basic Research
Development Program under Grant No. 2007CB815000, and the NSF of
China under Grants No. 10735010 and No. 10975006; and the UK STFC
and AWE plc.


\begin{thebibliography}{99}
\bibitem{WalNature99} P.M. Walker and G.D. Dracoulis,
Nature (London) {\bf 399}, 35 (1999).
\bibitem{WalHI01} P.M. Walker and G.D. Dracoulis,
Hyperfine Interact. {\bf 135}, 83 (2001).
\bibitem{WalPS83} P.M. Walker,
Phys. Scr. {\bf T5}, 29 (1983).
\bibitem{ReedPRL10} M.W. Reed {\it et al.}
Phys. Rev. Lett. {\bf 105}, 172501 (2010).
\bibitem{DraPLB06} G.D. Dracoulis {\it et al.},
Phys. Lett. B {\bf 635}, 200 (2006).
\bibitem{WalPT05} P.M. Walker and J.J. Carroll,
Phys. Today {\bf 58}, No. 6, 39 (2005).
\bibitem{JainNPA95} K. Jain {\it et al.},
Nucl. Phys. {\bf A591}, 61 (1995).
\bibitem{XuPRC00} F.R. Xu, P.M. Walker, and R. Wyss,
Phys. Rev. C {\bf 62}, 014301 (2000).
\bibitem{PodPLB00} Zs. Podoly\'{a}k {\it et al.},
Phys. Lett. B {\bf 491}, 225 (2000).
\bibitem{LaPRC10} G.J. Lane {\it et al.},
Phys. Rev. C {\bf 82}, 051304(R) (2010).
\bibitem{XuPLB98} F.R. Xu, P.M. Walker, J.A. Sheikh, and R. Wyss,
Phys. Lett. B {\bf 435}, 257 (1998).
\bibitem{LiuPRC11} H.L. Liu, F.R. Xu, P.M. Walker, and C.A. Bertulani,
Phys. Rev. C {\bf 83}, 011303(R) (2011).
\bibitem{NazNPA85} W. Nazarewicz, J. Dudek, R. Bengtsson, T. Bengtsson, and I. Ragnarsson,
Nucl. Phys. {\bf A435}, 397 (1985).
\bibitem{CwCPC87} S.\'{C}wiok, J. Dudek, W. Nazarewicz, S. Skalski, and T. Werner,
Comput. Phys. Commun. {\bf 46}, 379 (1987).
\bibitem{LipNPA73} H.C. Pradhan, Y. Nogami, and J. Law,
Nucl. Phys. {\bf A201}, 357 (1973).
\bibitem{MolNPA92} P. M{\"o}ller and J.R. Nix,
Nucl. Phys. {\bf A536}, 20 (1992).
\bibitem{MyeNP66} W.D. Myers and W.J. Swiatecki,
Nucl. Phys. {\bf 81}, 1 (1966).
\bibitem{StruNPA67} V.M. Strutinsky,
Nucl. Phys. {\bf A95}, 420 (1967).
\bibitem{AudNPA03} G. Audi, O. Bersillon, J. Blachot, and A.H. Wapstra,
Nucl. Phys. {\bf A729}, 3 (2003);
Evaluated Nuclear Structure Data File http:/www.nndc.bnl.gov/ensdf/.
\bibitem{MuPLB97} S.M. Mullins {\it et al.},
Phys. Lett. B {\bf 393}, 279 (1997); S.M. Mullins {\it et al.},
Phys. Lett. B {\bf 400}, 401 (1997).
\bibitem{PauPRC05} P.D. Stevenson, M.P. Brine, Zs. Podoly\'{a}k, P.H. Regan, P.M. Walker, and J.R. Stone,
Phys. Rev. C {\bf 72}, 047303 (2005).
\bibitem{SarPRC08} P. Sarriguren, R. Rodr\'{i}guez-Guzm\'{a}n, and L.M. Robledo,
Phys. Rev. C {\bf 77}, 064322 (2008).
\bibitem{Mo95} P. M{\"o}ller, J.R. Nix, W.D. Myers, and W.J. Swiatecki,
At. Data Nucl. Data Tables {\bf 59}, 185 (1995).
\bibitem{XuPRC99} F.R. Xu, P.M. Walker, and R. Wyss,
Phys. Rev. C {\bf 59}, 731 (1999).
\bibitem{JainRMP90} A.K. Jain, R.K. Sheline, P.C. Sood, and K. Jain,
Rev. Mod. Phys. {\bf 62}, 393 (1990).
\bibitem{WalPLB06} P.M. Walker and F.R. Xu,
Phys. Lett. B {\bf 635}, 286 (2006).
\bibitem{WalPLB97} P.M. Walker {\it et al.},
Phys. Lett. B {\bf 408}, 42 (1997).
\bibitem{NarNPA96} K. Narimatsu, R. Shimizu, and T. Shizuma,
Nucl. Phys. {\bf A601}, 69 (1996).

\end{thebibliography}
\end{document}